\documentclass[showpacs,preprint,amsmath,amssymb]{revtex4}
  \usepackage[dvips]{graphics}
\usepackage{graphicx}

\begin{document}

\title{Spin Photovoltaic Effect in Quantum Wires with Rashba Interaction}

\author{Yuriy V. Pershin}
\author{Carlo Piermarocchi}

\affiliation{\small Department of Physics and Astronomy, Michigan
State University, East Lansing, Michigan 48824-2320, USA}

\begin{abstract}
We propose a mechanism for spin polarized photocurrent generation
in quantum wires. The effect is due to the combined effect of
Rashba spin-orbit interaction, external magnetic field and
microwave radiation. The time-independent interactions in the wire
give rise to a spectrum asymmetry in $k$-space. The microwave
radiation induces transitions between spin-splitted subbands, and,
due to the peculiar energy dispersion relation, charge and spin
currents are generated at zero bias voltage. We demonstrate that
the generation of pure spin currents is possible under an
appropriate choice of external control parameters.
\end{abstract}


\maketitle

\newpage

The Rashba spin-orbit interaction (SOI) \cite{rashba} in transport
and equilibrium phenomena
\cite{jim1,streda,jim2,serra,governale,sayka,debald} plays a
central role in the fast growing fields of spintronics and quantum
computation \cite{zutic}. In particular, it was recently
discovered that the joint action of the Rashba SOI and in-plane
magnetic field on electrons confined in 1D quantum wires (QW)
results in unique properties \cite{jim1,streda,jim2}. Several
useful applications based on these properties were proposed,
including a scheme for measuring nuclear spin polarization
\cite{jim1} and a spin-filter \cite{streda}.

In this letter we theoretically investigate the effect of a
microwave radiation in a QW with Rashba SOI and an in-plane
magnetic field. This setup was stimulated by recent experiments on
the modifications of the Hall effect in the presence of a
microwave field \cite{mani}. We will show below that spin and
charge photocurrents can be generated in microwave irradiated QWs
(spin photovoltaic effect). The effect originates in the broken
symmetry of QW subbands caused by the interplay of SOI and
constant magnetic field. We emphasize that this mechanism is based
primarily on spin degrees of freedom in contrast to other
mechanisms of photovoltaic effect considered before (see, e.g.,
Ref. \cite{lenya}) and differs from the optical spin current
generation \cite{sipe}.

In our model a ballistic QW of length $L$ is connected to two
electron reservoirs having equal chemical potentials $\mu$. This
geometry is realized, for instance, in a QW created by a split
gate technique in a two-dimensional electron gas (2DEG). For the
sake of simplicity we assume that only the QW region is irradiated
by the microwave field. Without microwave radiation the currents
in QW from the left to the right reservoir and from the right to
the left reservoir balance each other so that the total current
through the QW is zero. The microwave radiation induces
transitions between spin-splitted subbands. The electron wave
vector $k$ is conserved in such transitions, however, in the
presence of SOI the electron velocity is not simply proportional
to $k$. The direction of the electron velocity, in specific
intervals of $k$, can be reversed after the transition. The
intersubband transition rate, due to the asymmetry of QW subbands,
is different for left- right-moving electrons, and produces a net
charge current. The spin current is also influenced by the
microwave field, since spin flips occur in these transitions.

We consider a QW in the $x$ direction created via a lateral
confinement (in $y$ direction) of a 2DEG in the ($x$, $y$) plane.
The Hamiltonian for the conduction electrons in the QW in the
presence of the microwave radiation can be written in the form
\cite{jim1,streda,jim2},
\begin{equation}
H=\frac{p^2}{2m^*}+V(y)-i\alpha \sigma_y \frac{\partial}{\partial
x} +\frac{g^*\mu_B}{2}\vec{\sigma}\cdot\vec{B}(t)+U(z,t).
\label{ham}
\end{equation}
We assume the microwave field propagating in the ($x$, $y$) plane
and we fix the electric field component in $z$ direction. In Eq.
(\ref{ham}), $\vec{p}$ is the momentum of the electron, $m^*$ is
the electron effective mass, $V(y)$ is the lateral confinement
potential due to the gates, $\alpha$ is the SOI constant,
$\vec{\sigma}$ is the vector of Pauli matrices, $\mu_B$ and $g^*$
are the Bohr magneton and effective g-factor, and $U(z,t)$ is the
potential due to the electric field component of the radiation.
$\vec{B}=\vec{B}_0+\vec{B}_1\cos(\omega t)$, where $\vec{B}_0$ is
the in-plane constant magnetic field and $\vec{B}_1$ is the
magnetic field component of the microwave radiation, and $\omega$
is the radiation frequency. $U(z,t)$ can be neglected because it
does not couple spin-splitted subbands. The third term in
(\ref{ham}) represents the Rashba SOI for an electron moving in
the $x$-direction. We assume that the effects of the Dresselhaus
SOI can be neglected \cite{ohno}.

At $B_1 = 0$, the solutions of the Schr\"odinger equation can be
written in the form
\begin{equation}
\Psi_{l,\pm}(k)=\frac{e^{ikx}}{\sqrt{2}}\binom{\pm
e^{i\varphi}}{1}\phi_l(y), \label{wf}
\end{equation}
where $\varphi=\textnormal{arctan}\left[-B_{0,y}/B_{0,x}-2\alpha k
/(g^* \mu_B B_{0,x}) \right]$ and $\phi_l(y)$ is the wave function
of the transverse modes (due to the confinement potential $V(y)$).
The eigenvalue problem can be solved to obtain
\begin{equation}
E_{l,\pm}(k)=\frac{E_Z^2}{E_\alpha}\tilde{k}^2+ E_l^{tr}\pm
E_Z\sqrt{1+\frac{2\tilde{k} B_{0,y}}{B_0}+\tilde{k}^2}.
\label{spectr}
\end{equation}
Here, $\tilde{k}=k\alpha/ E_Z$, $E_\alpha=2m^*\alpha^2/\hbar^2$,
$E_Z=g^*\mu_B B_0/2$, and $E_l^{tr}$ is the $l$-th eigenvalue of
$V(y)$. Assuming the parabolic confinement potential in the
$y$-direction, we have $E_l^{tr}=\hbar\omega_0 (l+1/2)$.

\begin{figure}[t]
\centering
\includegraphics[angle=270, width=12cm]{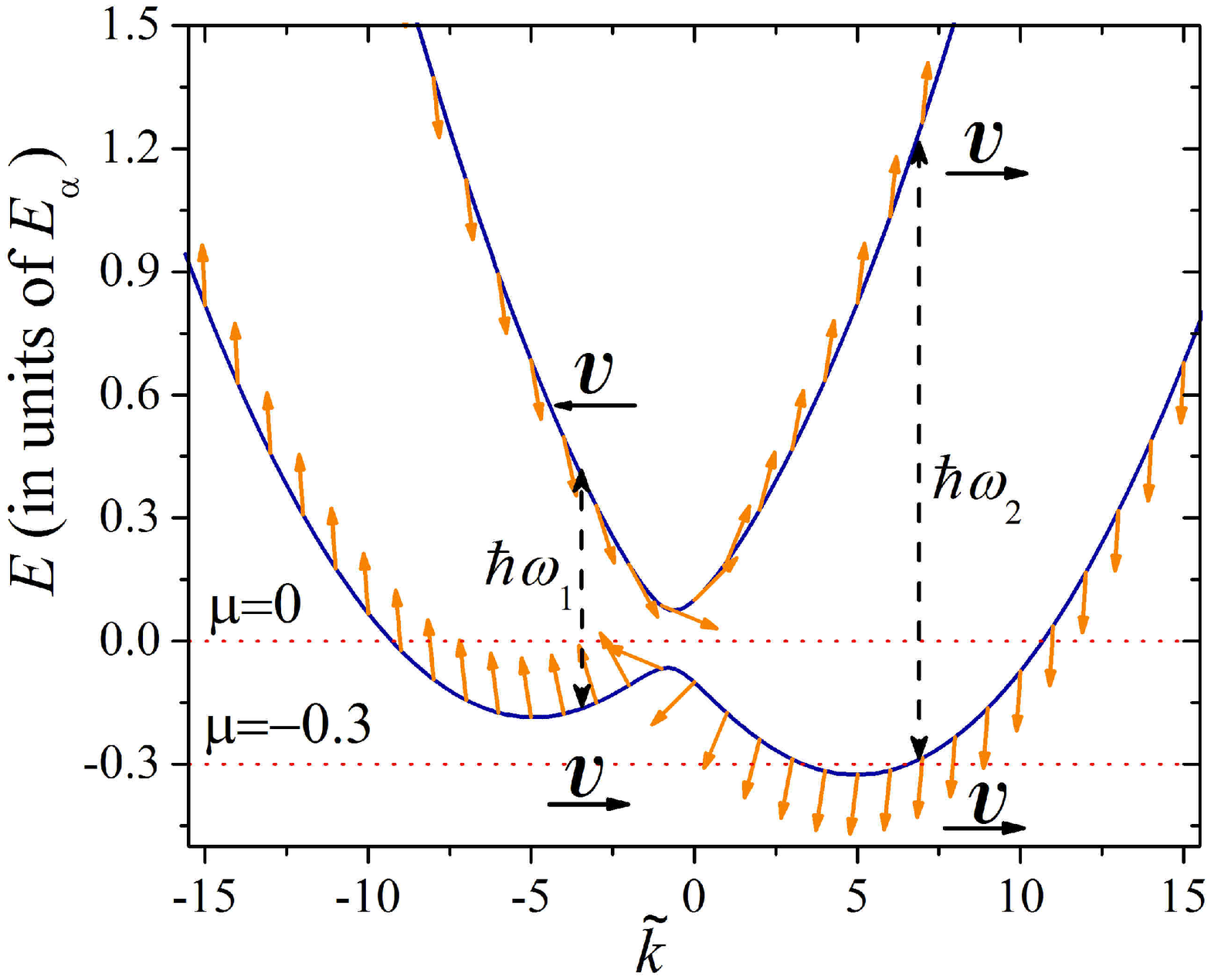}
\caption{} \label{fig1}
\end{figure}

The energy spectrum corresponding to Eq. (\ref{spectr}) is
illustrated in Fig. \ref{fig1}, for the two spin-split subbands
characterized by $l = 0$ and $\vec{B}_0$ directed at $\pi/4$ with
respect to the $x$-axis in ($x$, $y$) plane. Notice that the
down-splitted subband has a clearly defined asymmetry. This
subband features several local extrema, namely, two minima and one
maximum. The energy branches avoid the crossing and form a local
gap. The expectation values of spin polarization in the states
(\ref{wf}) are $\langle \pm|\sigma_x|\pm\rangle =\pm
\cos(\varphi(k))$ and $\langle \pm|\sigma_y|\pm\rangle =\mp
\sin(\varphi(k))$. While the external magnetic field realigns the
electron spins in the gap region, far from this region the spins
are polarized in $y$ direction by the Rashba SOI (Fig.
\ref{fig1}). The velocity of an electron is determined by
$v_{\pm}(k)=
\partial E_{\pm}/ \hbar \partial k$. Denoting by $k^{min}_+$, $k^{min,1}_-$ , $k^{max}_-$,
$k^{min,2}_-$ the positions of the local extrema of $E_{l,+}$ and
$E_{l,-}$, we find from Eq. (\ref{spectr}) that $v_+ < 0$ for
$k<k^{min}_+$, $v_+ > 0$ for $k>k^{min}_+$, $v_-<0$ for
$k<k^{min,1}_-$ and $k^{max}_-<k<k^{min,2}_-$, $v_- > 0$ for
$k^{min,1}_- < k < k^{max}_-$ and $k>k^{min,2}_-$. Generally,
$k^{min}_+ \neq k^{max}_-$. Thus, there are 2 intervals of $k$
when the directions of $v_-(k)$ and $v_+(k)$ are opposite.

Next we consider transitions generated by the time-dependent
magnetic field $\vec{B}_1 \cos(\omega t)$. It can be shown that
the transition rate differs from zero only for transitions between
subbands characterized by the same number $l$ with conservation of
$k$ (examples of such vertical transitions are presented in Fig.
\ref{fig1}). We calculate the transition rate using the expression
\begin{equation}
W=\frac{2\pi}{\hbar} \left| \left\langle + \left|\frac{g^* \mu_B
}{2} \vec{\sigma}\vec{B_1}\right|- \right\rangle \right|^2 \delta
\left( E_{l,+}-E_{l,-}-\hbar \omega \right). \label{matel}
\end{equation}
It follows from Eq. (\ref{matel}) that the transition rate depends
on the relative orientation of the spin polarization in a state
$k$ and the direction of $\vec{B}_1$. This property allows to use
the direction of the oscillating magnetic field as an additional
parameter that controls spin and charge currents in the QW.

\begin{figure}[b]
\centering
\includegraphics[angle=270, width=12cm]{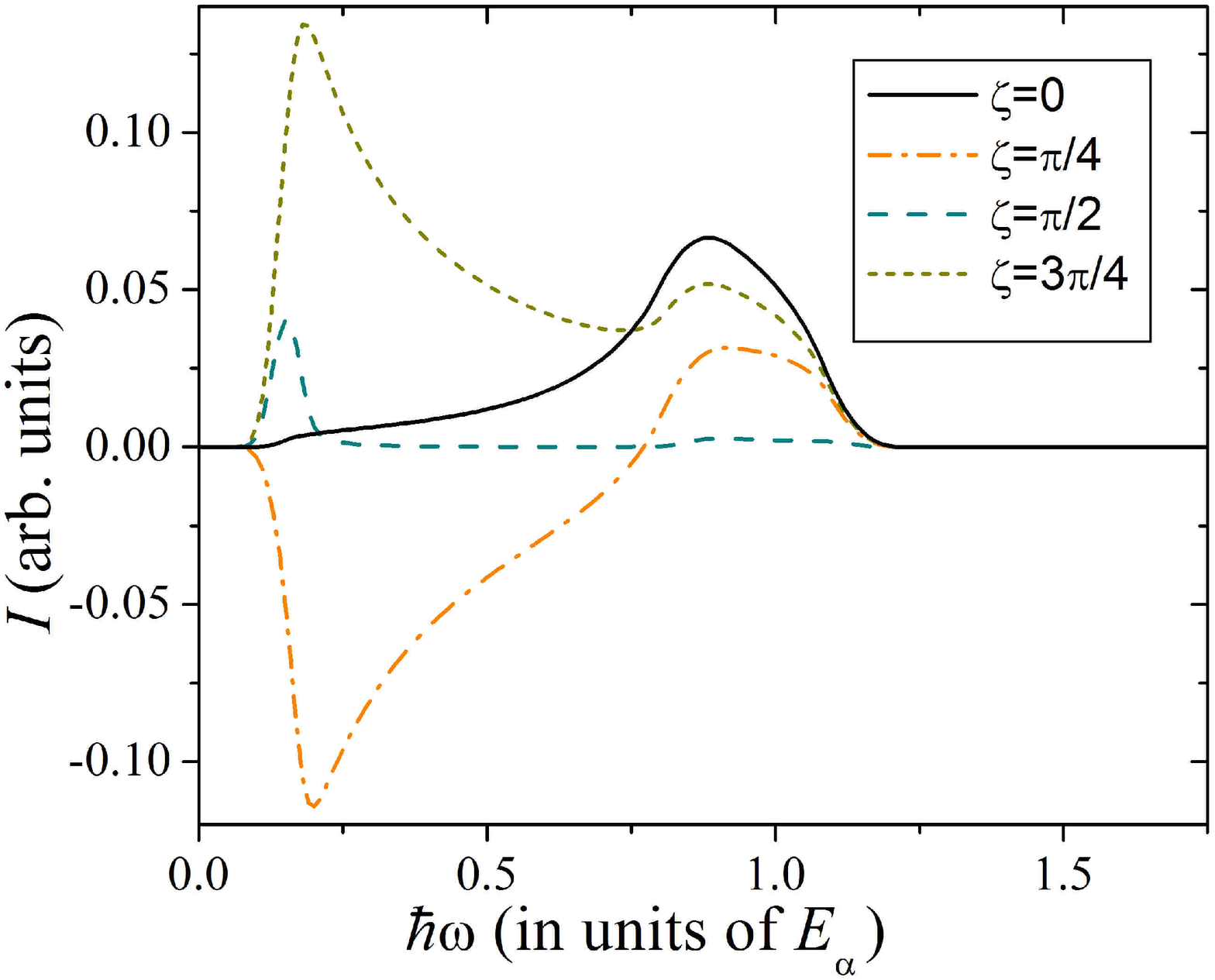}
\caption{} \label{fig2}
\end{figure}

In order to take into account the effect of the time-dependent
magnetic field on transport properties of the QW we solve the
Boltzmann equations
\begin{equation}
v_\mp(k)\frac{\partial f_\mp}{\partial
x}=Wf_\pm(1-f_\mp)-Wf_\mp(1-f_\pm), \label{boltz}
\end{equation}
 for the distribution functions $f_\pm(k, x)$. Assuming that the chemical
 potential is below the minimum of $E_+(k)$ and a low temperature,
 we supplement Eqs. (\ref{boltz}) by the following boundary conditions: for $v_-(k)
> 0$, $f_-(x = 0) = f(\mu)$; for $v_-(k) < 0$, $f_-(x = L) =
f(\mu)$; for $v_+(k)
> 0$, $f_+(x = 0) = 0$; for $v_+(k) < 0$, $f_+(x = L) = 0$, where
$f(\mu)$ is the Fermi function. The solution of linearized Eqs.
(\ref{boltz}) with the specified above boundary conditions was
found analytically. We calculate the current as \cite{datta}
$I=\frac{e}{h}\sum_{\nu=\pm} \int_{-\infty}^\infty v_\nu(k)
f_\nu(k)dk.$ We found that the transitions conserving the velocity
direction do not change the charge current. The current as a
function of the excitation frequency is shown in Fig. \ref{fig2}
and has a two-peak structure. The first peak in this structure
corresponds to transitions between the states with $k$ near
$k^{max}_-$, the second peak corresponds to transitions between
states with $k$ near $k^{min,2}_-$. The direction and amplitude of
the charge photocurrent shows a significant dependence on the
direction of $\vec{B}_1$, especially in the first peak region,
because the spin structure in this region is strongly affected by
the magnetic field direction.

\begin{figure}[bt]
\centering
\includegraphics[angle=270, width=12cm]{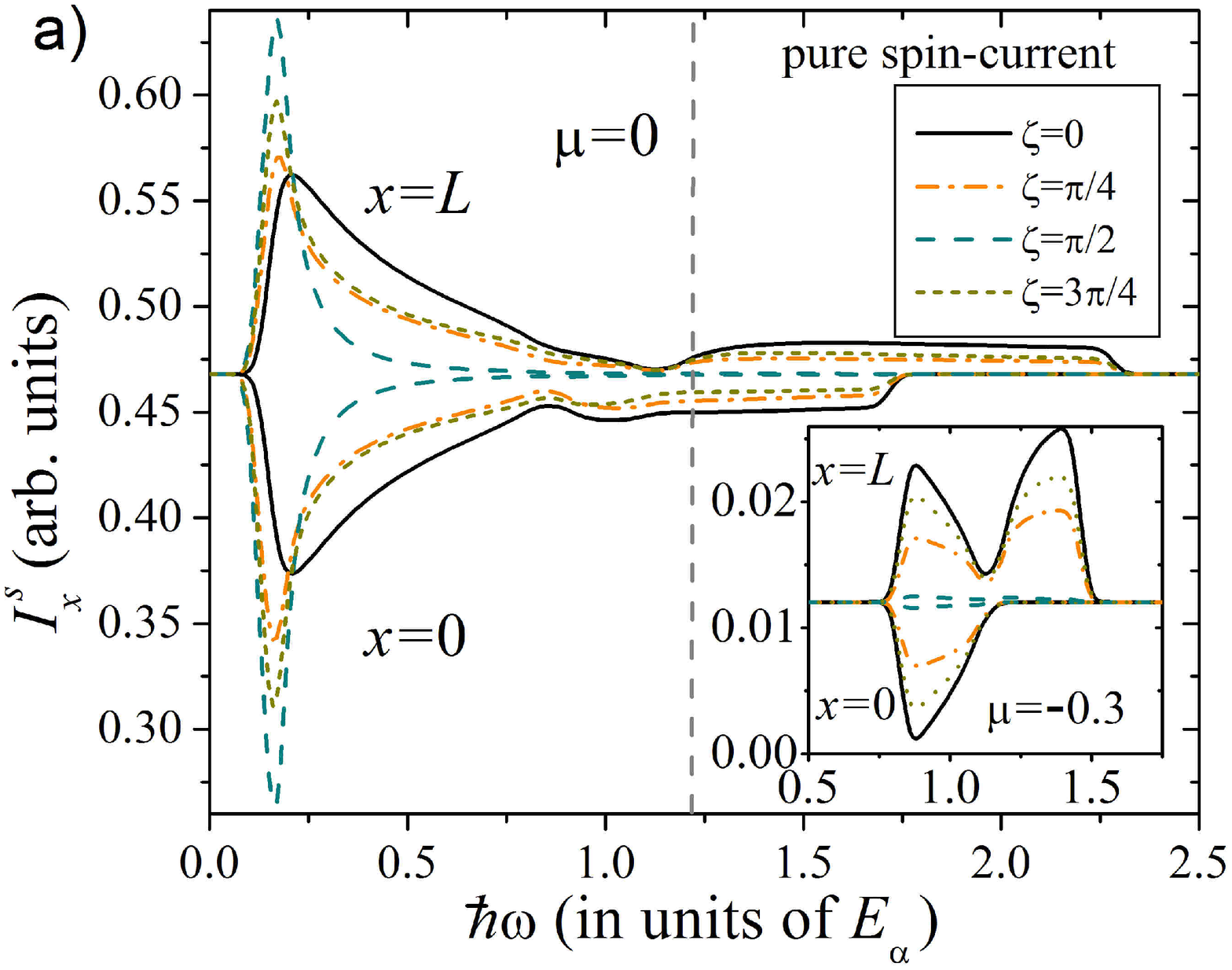}
\includegraphics[angle=270, width=12cm]{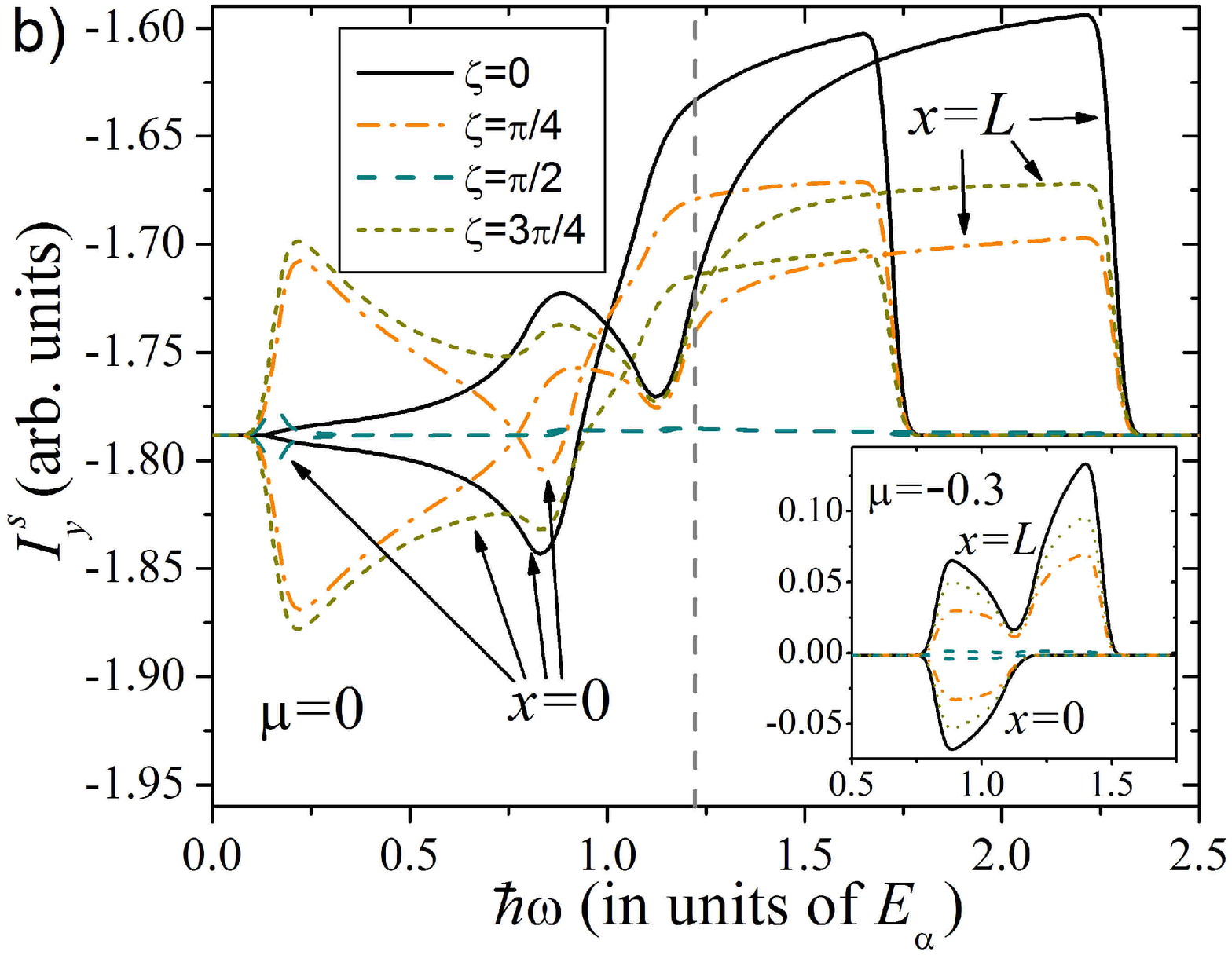}
\caption{} \label{fig3}
\end{figure}

As electrons carry spin as well as charge, the time-dependent
magnetic field also influences the spin current through the wire.
The spin current can be defined as the transport of electron spins
in real space. When the electron transport is confined to one
dimension, the spin current is a vector. Its components can be
calculated using $I^s_\gamma=\frac{1}{h}\sum_{\nu=\pm}
\int_{-\infty}^\infty \langle \nu|\sigma_\gamma|\nu\rangle
v_\nu(k) f_\nu(k)dk$ where $\gamma=(x,y,z)$. Fig. \ref{fig3} shows
the $x$ and $y$ components of the spin current ($I^s_z=0$)
calculated for two different values of the chemical potential
$\mu$. The main features of the spin current are: (i) the spin
current is coordinate-dependent; (ii) transitions conserving the
direction of the electron velocity contribute also to the spin
current; (iii) generation of a pure spin current (without a charge
current) occurs for transitions with $v_-(k)v_+(k)>0$. The spin
current components calculated at $\mu=0$ have a complex dependence
on $\omega$. At $\mu=-0.3E_\alpha$ the role of different
transitions can be more easily understood. With increase of
$\omega$, we first observe excitations with $v_-(k)v_+(k)<0$, and,
then, after passing the second minimum of $E_-(k)$ (see Fig.
\ref{fig1}), with $v_-(k)v_+(k)>0$. The insets in Fig. \ref{fig3}
show that the first type of transitions leads to changes of spin
current at $x=0$ and $x=L$, while the second type of transitions
changes the spin current at $x=L$ only. The asymmetry of the spin
current components at $x=0$ and $x=L$ is a signature of pure spin
currents in the system.

 We now discuss the conditions for an
experimental observation of this spin photovoltaic effect. First,
a QW should be fabricated from a structure with large Rashba SOI.
A promising candidate are  InAs based semiconductor
heterostructures, which have a relatively large $\alpha$
\cite{silsbee}. The characteristic energy of the SOI for these
structures ($\alpha = 4.5 \times 10^{-11}$eVm, $m^* = 0.036m_e$
\cite{Grundler}) is $E_\alpha = 1.9$meV. Assuming $E_Z =
0.1E_\alpha$ and taking $g^*= 6$ \cite{Grundler1} we obtain $B_0 =
1.1$T. Second, we note that extremely low temperatures are not
required for experimental observation of the spin photovoltaic
effect. From the condition $k_BT \lesssim E_\alpha$, $E_Z$ we
estimate $T \lesssim 2$K. Finally, the condition that $B_1 \neq 0$
only in the QW region was used only for convenience. In a real
experiment the whole system (QW and leads) is subjected to a
finite $B_1$. The effect of the leads depends on the particular
system studied, but should not considerably affect the scheme
proposed here, especially if there is no appreciable SOI in the
leads.

This research was supported by the National Science Foundation,
Grant NSF DMR-0312491.

\newpage

{ \bf Figure captions.}

\vspace{20 mm}

{ \bf Figure 1.} (Color online) Energy dispersion $E_{0,\pm}(k)$
(with respect to $E^{tr}_0$) for $E_Z = 0.1E_\alpha$ and
$\vec{B}_0 = (B_0/\sqrt{2},B_0/\sqrt{2}, 0)$. Spin orientation is
illustrated by arrows ($\langle\sigma_x\rangle$ is plotted along
$\tilde{k}$, $\langle\sigma_y\rangle$ is plotted along $E$, and
$\langle\sigma_z\rangle= 0$).

\vspace{10 mm}

{ \bf Figure 2.} (Color online) Current through QW as a function
of the excitation frequency $\omega$. $\zeta$ is the in-plane
angle between $\vec{B}_1$ and $x$ axis, $\mu = 0$, $T=0$.

\vspace{10 mm}

{ \bf Figure 3.} (Color online) Spin current components as a
function of the excitation frequency $\omega$, $\mu=0$, $T=0$. The
region of pure spin currents is to the right of the vertical
dashed line. Insets: spin current components at $\mu=-0.3$.

\newpage

\end{document}